\documentclass[12pt]{article}
\usepackage{a4}
\usepackage{epsf}
\usepackage{amssymb}
\usepackage{cite}
\newcommand{\be}{\begin{equation}}
\newcommand{\ee}{\end{equation}}
\newcommand{\bea}{\begin{eqnarray}}
\newcommand{\eea}{\end{eqnarray}}

\begin{document}
\def\C{{\mathbb{C}}}
\def\R{{\mathbb{R}}}
\def\s{{\mathbb{S}}}
\def\T{{\mathbb{T}}}
\def\Z{{\mathbb{Z}}}
\def\W{{\mathbb{W}}}
\def\Bbb{\mathbb}
\def\BZ{\Bbb Z} \def\BR{\Bbb R}
\def\BW{\Bbb W} 
\def\BM{\Bbb M} 
\def\e{\mbox{e}}
\def\BC{\Bbb C} \def\BP{\Bbb P}
\def\CP{\BC\BP}
\begin{titlepage}
\title{On Phases of Generic Toric Singularities}
\author{}
\date{Tapobrata Sarkar \thanks{\noindent tapo@iitk.ac.in},
Ajay Singh\thanks{\noindent sajay@iitk.ac.in}\\ 
\vspace{1.4cm}
Department of Physics, \\
Indian Institute of Technology,\\
Kanpur 208016, India}
\maketitle
\abstract{We systematically study the phases of generic toric
singularities, using methods initiated in hep-th/0612046. These 
correspond to Gauged Linear Sigma Models with arbitrary charges. 
We show that complete information about generic $U(1)^r$ GLSMs can 
be obtained by studying the GLSM Lagrangian, appropriately modified 
in the different phases of the theory. This can be used to study the 
different phases of $L^{a,b,c}$ spaces and their non-supersymmetric 
counterparts.  
}
\end{titlepage}

\section{Introduction}\label{intro}

In the last few years, there has been a lot of effort to understand 
the dynamics of spacetimes with non-trivial geometries in the 
framework of string theory. Several deep and fundamental insights have 
been obtained in the course of the study, which have remarkably 
contributed to our understanding of the underlying mathematical 
structure of singular spaces. The central tool in this study has been 
Witten's Gauged Linear Sigma Model \cite{wittenphases} (called GLSM in 
the sequel) a two dimensional $U(1)^r$ (world sheet) field theory, with
$(2,2)$ supersymmetry. It has by now been realised 
that the GLSM provides a very powerful tool in the analysis of 
stringy dynamics of non-trivial geometries, especially when these 
break space-time supersymmetry. A $U(1)^r$ GLSM describing a singular
space has $r$ Fayet-Iliopoulos parameters, which, for 
space-time non-supersymmetric theories are one loop renormalized. 
Tracking the flow of the GLSM under this RG gives us information about
the different phases of the theory. Indeed, a full understanding of 
these phases is essential in order to completely specify string theory 
on non-trivial backgrounds.  

Consider, e.g. a $U(1)$ GLSM of four chiral fields, with charges 
\begin{equation}
Q = \left(Q_1, Q_2, Q_3, -Q_4\right)
\label{charge1}
\end{equation}
When $\sum_i Q_i \neq 0$, the GLSM describes a non-supersymmetric
orbifold via the D-term equation
\begin{equation}
\sum_i Q_i |\phi_i|^2 + r = 0
\end{equation}
modulo the $U(1)$ identification. Varying the Fayet-Iliopoulos parameter
$r$ then determines the behaviour of the model at different points 
in the moduli space and gives us information about the various possible 
decays of this unstable orbifold under localised closed string tachyon 
condensation. Indeed, in order to fully understand stringy
dynamics in this model, one needs to specify the full set of possible
Fayet-Iliopoulos parameters in the theory (thereby enlarging the 
charge matrix) and in general this leads to a rich phase structure.
In particular, it has been shown that one can recover the complete 
set of D-brane charges in these theories by considering the 
Coulomb branch of the GLSM as well.

Apart from its utility in studying generic orbifold singularities, 
the GLSM has also been the central tool in the recent advances in 
our understanding of the extension of Maldacena's AdS/CFT  
correspondence \cite{malda}, involving $N=4$ Super Yang Mills theory,
to less supersymmetric situations. According to
its original formulation, the AdS/CFT correspondence states that
type IIB string theory on $AdS_5 \times S^5$, with appropriately
chosen R-R five form flux on the $S^5$ is dual to large N $N=4$ 
Super Yang-Mills theory. This duality has been refined since its  
inception to include more realistic situations with less supersymmetry,
and we now know that type IIB string theory on $AdS_5\times Y^5$, where
$Y^5$ is a Sasaki-Einstein manifold (i.e a manifold whose metric cone is
Calabi-Yau), with appropriate five form flux turned on, is dual to a 
four dimensional $N=1$ superconformal field theory (see, e.g 
\cite{klebwittetal}). Few explicit examples of Sasaki Einstein manifolds
were known till a few years back, when a major breakthrough was achieved 
in \cite{gmsw1}, where an infinite class of explicit Sasaki 
Einstein metrices with topology $S^2 \times S^3$ were constructed. 

Much work has followed since then, and the most general family of metrices
that have the topology of $S^2 \times S^3$ is denoted by $L^{a,b,c}$,
with $a, b, c$ being three positive integers. In the special case when
$a = p - q,~b=p+q,~c=p$, the $L^{a,b,c}$ metrics reduce to the family
of $Y^{p,q}$ metrics (see, e.g \cite{gmsw1}). As is well known, the dual  
$N=1$, $d=4$ SCFT in these cases naturally arises as the worldvolume 
low energy theory of a stack of D-3 branes probing a Calabi-Yau singularity, 
and residing at the tip of the singular Calabi-Yau cone. These are 
particularly simple yet illustrative examples, since the Calabi-Yau 
singularity is a toric variety. Indeed, the toric description of the
$Y^{p,q}$ class of metrics was provided in \cite{ms}, using which the
dual gauge theories were constructed in \cite{benvetal}. Further work
\cite{labctoric} has illustrated the GLSM approach to the more 
general $L^{a,b,c}$ spaces, and the main ingredient in the story is 
that the $L^{a,b,c}$ toric singularities arise as vacua of GLSMs, with 
charge matrices of the form 
\begin{equation}
Q = \left(Q_1, Q_2, -Q_3, -Q_4\right)
\end{equation}
where the $Q_i$s are positive (coprime) integers. This is in
distinction to the charge matrix in eq. (\ref{charge1}) where
three of the charges have the same sign. In applications to the
AdS/CFT correspondence, the charges are chosen such that they
sum to zero, in order to satisfy the Calabi-Yau condition, but
in general, this need not be true. When $\sum_iQ_i \neq 0$, the
GLSM describes an orbifold of the conifold singularity 
\cite{narayan1}, and results indicate that these might have, in
certain regions of moduli space, stable $L^{a,b,c}$ singularities.
In other words, as in case of orbifold theories, the singularity in 
question is unstable and decays to a stable singularity in the 
sense of the RG. Given the importance of generic $L^{a,b,c}$ spaces, 
it is especially important to understand the full phase structure of 
the GLSMs that may contain the latter in its phases. 

The phase structure mentioned above can be studied conveniently by 
constructing the GLSM Lagrangian in its most general form (i.e with
all the possible Fayet-Iliopoulos parameters turned on), and then 
tuning the various Fayet-Iliopoulos parameters 
of the theory to understand the various phases. The systematic 
procedure to do this was initiated in \cite{ts1}. In that paper, 
the generic $U(1)^r$ GLSM Lagrangian was constructed in the relevant
non-linear sigma model (NLSM) limit, and it was shown how the various 
phases of the GLSM corresponding to orbifold singularities can be studied 
in a very general fashion, by providing vevs to certain fields of the 
GLSM in accordance with the D-term equations of the model under 
consideration. Although the cases studied in \cite{ts1} were for 
orbifold singularities, these can be tuned to study more general GLSMs, 
like the supersymmetric examples considered in \cite{ms},\cite{labctoric}, 
or their non-supersymmetric counterparts \cite{narayan1}. Another 
aspect of the results in \cite{ts1} is that using the Lagrangian 
formulation of the GLSM in the NLSM limit, it is possible to study the 
behaviour of D-branes in various phases of the GLSM. In its simplest 
form, this problem reduces to constructing appropriate D-brane boundary 
conditions in the GLSM \cite{hiv}, \cite{gjs} and then continuing 
these appropriately to the different regions of the GLSM moduli space.  

The purpose of the present paper is to use and extend the ideas developed 
in \cite{ts1} to analyse, in most general terms, the phase structure of 
generic GLSMs corresponding to unstable spaces. 
We will show in the course of this paper that the phases
of arbitrary charged GLSMs can be analysed in a fully algebraic
approach, which makes our methods computationally simpler than other
existing techniques. We work out several examples as an illustration
of our approach. The organisation of the paper is as follows. In 
section 2, we review some basic results on the GLSM Lagrangians that 
were obtained in \cite{ts1}. In section 3, which is the main part of the
paper, we use this Lagrangian formulation to study and extend the 
analysis of \cite{ts1} to analyse the phases of GLSMs describing arbitrary 
toric singularities.  Finally, section 4 concludes with some discussions,     
and possible extensions of our work.

\section{The GLSM Lagrangian and Singular Spaces}

In this section, we briefly review the results of \cite{ts1} in 
analysing the phases of GLSMs corresponding to generic orbifold 
singularities. This section is review material, and is meant 
to set the notations and conventions to be used in the rest of the paper.

Since much of what follows in this paper deals with unstable spaces,
let us begin by briefly reviewing the notion of the simplest 
types of unstable spaces which can be given a toric description, i.e
the non-supersymmetric orbifolds of $\BC^2$ and their decay properties. 
Consider the orbifold of $\BC^2$, with action 
\begin{equation}
\left(Z_1, Z_2\right) \to \left(\omega Z_1, \omega^p Z_2\right)
\end{equation}
where $Z_1$ and $Z_2$ are coordinates on the $\BC^2$, and 
$\omega = e^{\frac{2\pi i}{n}}$ is the $n$ th root of unity.
When $p \neq n-1$, this orbifold action breaks space-time supersymmetry,
and introduces tachyons in the closed string spectrum for both 
Type II and Type 0 strings, that are localised at the tip of the
orbifold (and can be interpreted as twisted sector states in the
closed string worlds sheet conformal field theory). A similar action
can be written down for $\BC^3$ (or $\BC$) orbifolds.

Condensation of closed string tachyons then shows the non-supersymmetric 
orbifolds decays toward more stable configurations. Whereas for orbifolds 
of $\BC$, the end product of the decay is always flat space, orbifolds
of $\BC^2$ and $\BC^3$ show a much richer structure. Whereas the 
end product of decay of $\BC^2$ orbifolds are generally supersymmetric
orbifolds of lower rank, for $\BC^3$ orbifolds, one might end up 
reaching a terminal singularity.   
 
The ``brane probe'' approach of Adams, Polchinski and Silverstein 
(APS) \cite{aps} who first studied these singularities was to use a 
probe D-brane that has its world volume transverse to the orbifolded 
directions and is stuck and the orbifold fixed point. The brane probe 
picture is essentially an open string picture in the substringy regime with 
localised tachyons, and can be studied by using the gauge theory
living on the world volume of the D-brane.  In the APS procedure, 
it is found that by exciting the marginal or tachyonic deformations in 
the theory, one can drive the original orbifold to one of the lower 
rank and possible tachyonic deformations of the latter takes the system to a
final stable supersymmetric configuration. An useful alternative 
approach is to study the $N=(2,2)$ CFT of the worldsheet 
which is related to Witten's GLSM. \cite{vafa}. One can construct an 
appropriate GLSM corresponding to the non-supersymmetric orbifolds, 
and track the behaviour of the model in the sense of the RG, and this 
effectively describes the decays of these orbifolds. The GLSM has a very 
rich phase structure that can be studied by tuning 
the relevant Fayet-Iliopoulos parameters that appear in the theory.
 
In a different approach to the problem of tachyon condensation, 
the sigma model metrics (with multiple $U(1)$ gauge groups) 
was calculated for the non-supersymmetric $\BC^2/\BZ_n$ and 
$\BC^3/\BZ_n$ orbifolds \cite{ts2},\cite{mt}. The advantage of this 
method is that it can be used to study the phases of generic GLSMs, 
without having to resort to a case by case analysis. Also, D-brane 
dynamics in generic orbifolds of $\BC^r$ can be understood in terms of 
open string GLSM boundary conditions \cite{hiv},\cite{ts1}.

In order to illustrate the above, let us begin with a brief description 
of the GLSM. The action for a GLSM with, with an Abelian gauge group 
$U(1)$ is given by

\begin{equation} 
S = \int d^2z d^4\theta \sum_i \bar{\Phi}_i \Phi_i - 
\sum \frac{1}{4e^2} \int d^2 z d^4\theta \bar{\Sigma}_a \Sigma_a 
+ Re \left[ it \int d^2 z d^2 \tilde{\theta} \Sigma \right] 
\label{action}
\end{equation}
where the $\Phi_i$ are chiral superfields, $\Sigma_a $ is a twisted 
chiral superfield, $t= ir + \frac{\theta}{2\pi}$ 
is a complexified parameter involving the Fayet-Iliopoulos parameter 
$r$ and the two dimensional $ \theta $ angle. As appropriate
in our case, we consider a theory without a superpotential. In 
general, we will consider GLSMs with multiple $U(1)$ gauge groups. 

\noindent
In the $e^2\rightarrow \infty$ limit of the GLSM, the gauge fields 
appearing in (\ref{action}) are Lagrange multipliers. It is 
then possible to obtain the Lagrangian and solve the D-term constraint 
in the classical limit $|r| \rightarrow \infty $ to read off the sigma 
model metric corresponding to the GLSM \cite{mt},\cite{ts1}. 
Focusing on the bosonic part of the action in (\ref{action}), given by
\begin{equation} 
S = - \int d^2z D_\mu \bar{\phi}_i D^\mu \phi_i 
\label{action1}
\end{equation}
the Lagrangian can be studied using the D-term constraints,
\begin{equation}
\sum_i Q_i^a |\phi_i|^2 + r_a = 0
\end{equation}
where $\phi_i$ are the bosonic components of the $\Phi_i$ and $Q_i^a$ 
denote the charges of the $\phi_i$ with respect to the $a$th $U(1)$.
Orbifolds of the type $\BC^r/\Gamma$, with $r=1,2,3$ can be 
described by GLSM, with the number of the gauge groups being dictated 
by the nature of the singularity. In the NLSM limit, the component 
gauge fields in the model can be calculated and substituted back into 
the action to get the GLSM Lagrangian entirely in terms of the
toric data of the singularity. 

The Lagrangian for a GLSM with the $m$ fields $\phi_i$, $i=1,2,...,m$ 
with single $U(1)$ gauge group with charges $Q_i$, $i=1,2,...m$ 
is given by:
\begin{equation}
L = (\partial_{\mu}{\rho_1})^2 + (\partial_{\mu}{\rho_2})^2 + \cdots 
+(\partial_{\mu}{\rho_m})^2 +\frac{\sum_{i<j}{\rho_i^2}{\rho_j^2}
(Q_i \partial_{\mu}{\theta_j}-Q_j \partial_{\mu}{\theta_i})^2}
{\sum_j Q^2_j \rho_j^2}
\label{onepar}
\end{equation}
In the classical limits of the Fayet-Iliopoulos parameter, this 
formula gives the sigma model metric for the 
singularity $\BC^{m-1}/\BZ_n$. As we have mentioned before, this 
corresponds to giving a large vev to any of the fields appearing in 
the Lagrangian.  

Following a similar approach, the Lagrangian for the two parameter 
GLSM can be constructed. For $m$ fields $\phi_i$, $i=1,2,...,m$ and two 
gauge groups $a,b=1,2$, the expression for the Lagrangian is:
\begin{equation}
L = L_1 + L_2
\label{twopar1}
\end{equation}
where
\begin{equation}
L_1 = \sum_i(\partial_{\mu}\rho_i)^2
\label{twopar2}
\end{equation}
\begin{equation}
L_2 = \frac{\sum_{[i,j,k]}[\rho_i \rho_j \rho_k \partial_{\mu}
\theta_i(Q^b_j Q^a_k- Q^b_k Q^a_j)]^2}{\sum_{i<j} \rho_i^2 
\rho_j^2 (Q^b_i Q^a_j- Q^a_i Q^b_j)}
\label{twopar3}
\end{equation}
Where the symbol $[i,j,k]$ in the summation in the numerator in $L_2$ 
denotes $cyclic$ combinations of the variables and we have written 
$\phi_i =\rho_i e^{i\theta_i}$. This expression can be used to study 
non-cyclic singularities of the form $\BC^3 / \BZ_m \times \BZ_n $, 
which can not be described by the single parameter GLSM.

The above Lagrangians can be generalized to the case of the general 
$r$ parameter GLSMs. The general $r$ parameter GLSM 
Lagrangian can be written as 
\begin{equation}
L = L_1+L_2
\label{rpar1}
\end{equation}
where now
\begin{equation}
L_1= \sum_i (\rho_i)^2
\label{rpar2}
\end{equation}
\begin{equation}
L_2 = \frac{\sum_{[j_1,j_2,...,j_{r+1}]} [\rho_{j_1}\rho_{j_2}...
\rho_{j_r} \partial_{\mu}(\theta_{j_1} K_{j_2,...,j_r})]^2}
{\sum_{j_1<j_2<...<j_r}\rho_{j_1}^2\rho_{j_2}^2...\rho_{j_r}^2 
[\Delta(j_1,j_2,...,j_r)]^2}
\label{rpar3}
\end{equation}
where $i$ and $j_1,j_2,...,j_{r+1}$ go from $1,2,...,n$, where $n$ is 
the total number of scalar fields. $K_{j_2,...,j_r}$ is the $j_1th$ 
component of the kernel of the matrix formed by the charges of the 
$j_{r+1}$ vectors in the numerator of $L_2$(and hence depends on 
$j_2,j_3,...,j_{r+1}$), and $\Delta(j_1,j_2,...,j_r)$ is the determinant 
of the matrix formed by the charge vectors 
$\rho_{j_1},\rho_{j_2},...,\rho_{j_r}$ under the $r$ $U(1)s$. Again the 
notation $[j_1,j_2,...,j_{r+1}]$ indicates a cyclic combination of the 
variables.

Having written down the GLSM Lagrangian in its most general form entirely 
in terms of the toric data of the orbifold, this formalism can be used to 
study the phases of arbitrary charge GLSMs. For a multi-parameter 
GLSM, these phases are obtained from the Lagrangian by making some of 
the fields in the GLSM very large. \footnote{Strictly speaking, this 
corresponds to a region of the moduli space where some combination 
of the Fayet-Iliopoulos parameters being very large.} In \cite{ts1}, 
this approach was used to study the phases of orbifold GLSMs, and it 
was shown how non-cyclic orbifolds of $\BC^3$, i.e orbifolds of the form
$\BC^3/\BZ_n\times\BZ_m$ can be handled easily in this formalism.  
In particular, \cite{ts1} dealt with the computation of the sigma model
metrics in the various phases of the GLSM. In the next section, we extend 
these results of \cite{ts1} and study the phases of arbitrary GLSMs. 

\section{GLSM Analysis of Generic Singular Spaces}

In this section, we will study the phases of arbitrary GLSMs, extending
the analysis of \cite{ts1}, using the Lagrangian formulation developed
therein and discussed in the previous section. Whereas our previous 
study focused on the sigma model metrics in phases of orbifold 
singularities, we will be more general here. Generically, given 
an $r$ parameter GLSM with say $m$ fields, we would like to see the 
effect of giving vevs to an arbitrary number of fields. 

To begin with, note that there is a subtlety involved in our
Lagrangian formulation. Consider, e.g. the two parameter GLSM 
of eqs. (\ref{twopar3}) and (\ref{rpar3}). Namely, in these 
equations, the square of the charges appear
in the denominator, and hence when some of the fields are set to be 
very large, it might seem that there is a sign ambiguity in 
the definition of the charges of the remaining fields in the reduced 
Lagrangian.  However, it is not difficult to convince oneself that there 
is actually no such ambiguity. It is best to illustrate this with an 
example.  Consider, e.g. the GLSM corresponding to the unstable 
orbifold $\BC^2/\BZ_{5(3)}$. The closed string description of
this singularity tells us that there are two twisted sectors that 
participate in the full resolution (corresponding to divisors with 
intersection numbers $-2$ and $-3$, and hence the $U(1)^2$ charge 
matrix for this singularity is given by
\begin{equation}
Q = \pmatrix{1&3&-5&0 \cr 2&1&0&-5}
\label{charge53}
\end{equation}
Writing the fields as $\phi_i = \rho_ie^{i\theta_i}$, in the limit 
that one of the fields, say $|\phi_1| \gg 0$, we substitute
this charge matrix in eq. (\ref{twopar2}), to obtain the (relevant
part of the) reduced Lagrangian $L = \frac{N}{D}$ where now
\begin{eqnarray}
N &=& \rho_2^2\rho_3^2\left(-\partial_{\mu}\theta_3 -2\partial_{\mu}\theta_2
+ \partial_{\mu}\theta_1\right)^2 + 
\rho_2^2\rho_4^2\left(-\partial_{\mu}\theta_4 +\partial_{\mu}\theta_2
- 3\partial_{\mu}\theta_1\right)^2 \nonumber\\
&+& \rho_3^2\rho_4^2\left(2\partial_{\mu}\theta_4 +\partial_{\mu}\theta_3
+ 5\partial_{\mu}\theta_1\right)^2 \nonumber\\
D &=& \left[\rho_2^2\left(1\right)^2 + \rho_3^2\left(-2\right)^2
+ \rho_4^2\left(+1\right)^2\right]
\label{53a}
\end{eqnarray}
Note that the terms in $N$ correspond to the gauge invariant angles, 
and we have explicitly indicated the fact that in the denominator, 
the original signs appearing with the various terms in eq. (\ref{twopar2}) 
have to be retained (modulo possibly and overall relative sign between 
the terms). The value of $D$ shows that we now have a reduced charge 
matrix for the fields $\phi_2, \phi_3, \phi_4$ with 
\begin{equation}
Q = \left(1,-2,1\right)
\end{equation}
which is the GLSM for the supersymmetric orbifold $\BC^2/\BZ_2$.
Let us make a few comments at this stage. In general, in a two parameter
GLSM, making one field large (i.e giving it a large vev) will not break
the full $U(1)^2$ symmetry. Consider, e.g. the charge matrix in 
eq. (\ref{charge53}). The two D-term constraints coming from this
charge matrix is given by
\begin{eqnarray}
|\phi_1|^2 + 3|\phi_2|^2 - 5|\phi_3|^2 + r_1 = 0 \nonumber\\
2|\phi_1|^2 + |\phi_2|^2 - 5|\phi_4|^2 + r_2 = 0 \nonumber\\
\end{eqnarray}
Setting $r_1 \ll 0$, we can solve for $|\phi_1|$ as
\begin{equation}
|\phi_1| = \sqrt{5|\phi_3|^2 - 3|\phi_2|^2 - r_1}
\end{equation}
Now, substituting this value of $|\phi_1|$ in the second of the
D-term equations, we see that there is a residual unbroken $U(1)$
with a modified D-term constraint
\begin{equation}
-5|\phi_2|^2 + 10|\phi_3|^2 -5|\phi_4|^2 + \left(r_2 - 2r_1\right) = 0
\end{equation}
In order to completely break the original $U(1)^2$, we now need to give
a vev to a second field. Hence, the GLSM that we obtain by making one
field very large refers to this residual $U(1)$. 
Now, the relevant part of the one parameter Lagrangian 
\footnote{In the $L_1$ component in eq. (\ref{twopar2}) or eq. 
(\ref{rpar2}), the field that has been made large drops out} 
with this charge matrix is given by
$L = N_1/D_1$, where $D_1 = D$ of eq. (\ref{53a}) and 
\begin{equation}
N_1 = \rho_2^2\rho_3^2\left(\partial_{\mu}{\tilde \theta_3} 
+ 2 \partial_{\mu}{\tilde \theta_2}\right)^2 +
\rho_2^2\rho_4^2\left(\partial_{\mu}{\tilde \theta_4}
- \partial_{\mu}{\tilde \theta_2}\right)^2 +
\rho_3^2\rho_4^2\left(-2\partial_{\mu}{\tilde \theta_4}
- \partial_{\mu}{\tilde \theta_3}\right)^2  
\end{equation}
where we have denoted the angular variables in the reduced Lagrangian
with a tilde. Now, with the identification
\begin{equation}
{\tilde \theta_2} = \theta_2,~~
{\tilde \theta_3} = \theta_3 - \theta_1,~~
{\tilde \theta_4} = \theta_4 + 3\theta_1
\end{equation}
we see that the two Lagrangians are identical. This analysis tells us
that the supersymmetric $\BC^2/\BZ_2$ orbifold arises as a decay product 
of the unstable $\BC^2/\BZ_{5(3)}$ orbifold. 
Given the generality of our analysis, it should be clear that this 
can be used to analyse the phases of {\it any} GLSM with an arbitrary 
number of gauge groups and arbitrary charges. 

With these comments, We will now begin our analysis of singular 
spaces corresponding to GLSMs with charges 
\begin{equation}
Q = \left(Q_1, Q_2, -Q_3, -Q_4\right)
\label{charges}
\end{equation}
with the $Q_i$ being positive integers. As discussed in 
\cite{benvetal}, this is the most general charge configuration 
for a $U(1)$ GLSM with four fields which does not describe an orbifold 
singularity. This is becos all the charges have been taken to be non-zero, 
hence either two of them or three of them have the same sign, but the 
latter are simply orbifolds of $\BC^3$ so for our purposes, it is enough 
to begin with the charge matrix of eq. (\ref{charges}). For the Calabi-Yau 
condition to be satisfied, one requires that $\sum_i Q_i = 0$, but 
we will not put such a restriction here, and would consider the
general case where $\sum_i Q_i \neq 0$. The Lagrangian
corresponding to the infinite gauge coupling limit of the GLSM, with 
the D-term constraint being
\begin{equation}
Q_1|\phi_1|^2 + Q_2|\phi_2|^2 - Q_3|\phi_3|^2 - Q_4|\phi_4|^2 +r = 0
\end{equation}  
is given by setting $m=4$ in eq. (\ref{onepar}), 
\begin{equation}
L = (\partial_{\mu}{\rho_1})^2 + \cdots +(\partial_{\mu}{\rho_4})^2 
+\frac{\sum_{i,j=1,\cdots 4,i<j}{\rho_i^2}{\rho_j^2}
(Q_i \partial_{\mu}{\theta_j}-
Q_j \partial_{\mu}{\theta_i})^2}{\sum_j Q^2_j \rho_j^2}
\label{lag4charge}
\end{equation}
where the fields with charge $Q_i$ have been written as 
$\phi_i = \rho_ie^{i\theta_i}$. 

We now look at the classical limits of this GLSM. This can be done by
setting the (magnitude of the) Fayet-Iliopoulos parameter to be very large.
Specifically, setting $r$ to be very large positive, we see that either
$\phi_3$ or $\phi_4$ has to be made very large. \footnote{Equivalently,
both these fields can be made very large, as we will see in a moment.} 
If we choose $\phi_4$ to be very large, we can solve for the fields in 
the classical limit as 
\begin{equation}
\phi_1 = \rho_1e^{i\theta_1},~~
\phi_2 = \rho_2e^{i\theta_2},~~
\phi_3 = \rho_3e^{i\theta_3},~~
\phi_4 = \sqrt{\frac{Q_1\rho_1^2 + Q_2\rho_2^2 - Q_3\rho_3^2 + r}{Q_4}}
e^{i\theta_4}
\end{equation}
Substituting these values in the Lagrangian yields 
\begin{equation}
L = \sum_{i=1}^{3}\left(\partial_{\mu}\rho_i\right)^2 
+ \rho_1^2d{\tilde\theta_1}^2 + \rho_2^2d{\tilde\theta_2}^2
+ \rho_3^2d{\tilde\theta_3}^2
\label{laglabc1}
\end{equation}
where
\begin{equation}
{\tilde\theta_1} = \theta_1 + \frac{Q_1}{Q_4}\theta_4,~~
{\tilde\theta_2} = \theta_2 + \frac{Q_2}{Q_4}\theta_4,~~
{\tilde\theta_3} = \theta_3 - \frac{Q_3}{Q_4}\theta_4
\label{laglabc2}
\end{equation}
This can be recognised as the Lagrangian corresponding to the orbifold
GLSM with charges 
\begin{equation}
Q = \left(Q_1, Q_2, pQ_4 - Q_3, -Q_4\right)
\end{equation}
where $p$ is the smallest positive integer that makes $pQ_4 - Q_3$ a 
positive number. Similarly, if we set $\phi_3$ to be very large, we 
obtain the Lagrangian corresponding to the classical limit of the GLSM 
with charges
\begin{equation}
Q = \left(Q_1, Q_2, p'Q_3 - Q_4, -Q_3\right)
\end{equation}
where, as before we have introduced an integer $p'$ to make the third entry
in the above equation positive.

The analysis for $r \ll 0$ can be carried out in exactly the same way,
and the corresponding orbifold singularities have ranks $Q_1$ and
$Q_2$. This shows that the GLSM with charges given in eq. (\ref{charges})
contain orbifold singularities in their classical limits (This conclusion
has been reached by other methods in \cite{narayan1}). The full phase
structure of the GLSM can thus be studied by including the additional
blow up modes that follow from these orbifolds. Let us see if we can
substantiate this. Consider, e.g. the simpler class of the supersymmetric 
$Y^{p,q}$ singularities, described by the GLSM with charge matrix
\begin{equation}
Q = \left(p-q, p+q, -p, -p\right)
\label{chargeypq}
\end{equation}
The D-term constraint in this case reads
\begin{equation}
\left(p-q\right)|\phi_1|^2 + \left(p+q\right)|\phi_2|^2 
- p\left(|\phi_3|^2 + |\phi_4|^2\right) + r = 0
\label{dtermypq}
\end{equation}
in the classical limits, we can solve the D-term constraint as before.
Consider e.g. the limit $r \gg 0$. In this case, we can choose to set
the magnitude of $\phi_4$ to be very large. Substituting the result in
eq. (\ref{lag4charge}) we see that the resulting Lagrangian has the same 
form as that in eqs. (\ref{laglabc1}) and (\ref{laglabc2}), excepting
that now the coordinate corresponding to $\rho_3$ (and $\theta_3$) are
unorbifolded, leading to the fact that in this limit we actually
have a supersymmetric $\BC^2/\BZ_p$ singularity. A similar result is
obtained on setting  $\phi_3 \gg 0$ wherein we recover the same 
singularity. In the other limit, i.e when $r \ll 0$, we recover two
supersymmetric $\BC^3$ orbifolds, of ranks $p-q$ and $p+q$. Let us
take the concrete example of the GLSM corresponding to $Y^{3,2}$, 
given by the charge matrix
\begin{equation}
Q = \left(1,5, -3, -3\right)
\end{equation}
The discussion in the preceding paragraph tells us that in the
various classical limits of the Fayet-Iliopoulos parameter of this
model, we recover, apart from flat space, the $\BC^3/\BZ_5$ orbifold
and two copies of the orbifold $\BC^2/\BZ_3 \times \BC$
\footnote{The action of the $\BC^3$ orbifold is
$\left(Z_1, Z_2, Z_3\right) \to \left(\omega Z_1, \omega^2Z_2,
\omega^2Z_3\right)$ with $\omega = e^{\frac{2\pi i}{5}}$ 
and that of the $\BC^2$ orbifolds is 
$\left(Z_1, Z_2\right) \to \left(\omega' Z_1, \omega'^2Z_2\right)$
with $\omega' = e^{\frac{2\pi i}{3}}$}
The original GLSM charge matrix can now be enhanced by adding the 
twisted sectors corresponding to marginal deformations, and the full
GLSM charge matrix is calculated to be
\begin{equation}
Q = \pmatrix{1&5&-3&-3&0&0&0&0\cr1&0&2&2&-5&0&0&0\cr
3&0&1&1&0&-5&0&0\cr 1&2&0&0&0&0&-3&0\cr 2&1&0&0&0&0&0&-3}
\label{fullchargey32}
\end{equation}
The complete phase structure of the $Y^{3,2}$ space can now be obtained
by analysing the Lagrangian corresponding to the charges of 
eq. (\ref{fullchargey32}) by making any combination of fields very
large. Since the theory is supersymmetric, all the added twisted
sector charges survive the GSO projection. This will in general not
be the case for unstable spaces. For the charge matrix of 
eq. (\ref{fullchargey32}), we present the results for some of the
phases of the theory. E.g if we make the fields $\rho_1,\rho_2,\rho_3,
\rho_5$ and $\rho_6$ very large, the resultant flat sigma model metric is
\begin{eqnarray}
ds^2 &=& d\rho_4^2 + d\rho_7^2 + d\rho_8^2 + \rho_4^2
d(\theta_4-\theta_3)^2 + \rho_7^2 d(\theta_7-\theta_1+2\theta_2
+3\theta_3+\theta_5)^2 +
\nonumber \\
&& \rho_8^2d(\theta_8+2\theta_1-\theta_2-\theta_3+\theta_6)^2
\end{eqnarray}
Making the fields $\rho_2,\rho_4,\rho_6,\rho_7$ and $\rho_8$ acquire
very large vevs, we obtain the sigma model metric
\begin{eqnarray}
ds^2 &=& d\rho_1^2 + d\rho_3^2 + d\rho_5^2 +
\frac{\rho_1^2}{(2)^2}d(2\theta_1-\theta_2-\theta_4+\theta_6+\theta_8)^2 +
\rho_3^2d(\theta_3-\theta_4)^2 +
\nonumber \\
&& \frac{\rho_5^2}{(2)^2}d(2\theta_5+3\theta_2+5\theta_4+\theta_6
+2\theta_7+\theta_8)^2
\end{eqnarray}
which can be recognised to be the metric for $\BC^2/\BZ_2 \times \BC$, 
and arises in a limit of the supersymmetric $\BC^2/\BZ_3 \times \BC$ 
that we have seen earlier. Finally, say we look at the region of moduli
space where the fields $\rho_1,\rho_2,\rho_5,\rho_6$ and $\rho_7$ acquire
large vevs. In that case, the metric reads
\begin{eqnarray}
ds^2 &=& d\rho_3^2 + d\rho_4^2 + d\rho_8^2 +
\frac{\rho_3^2}{(3)^2}d(3\theta_3-\theta_1+2\theta_2+\theta_5+\theta_7)^2 +
\nonumber \\
&&\frac{\rho_4^2}{(3)^2}d(3\theta_4-\theta_1+2\theta_2+\theta_5+\theta_7)^2 +
\nonumber \\
&& \frac{\rho_8^2}{(3)^2} d(3\theta_8+5\theta_1-\theta_2+\theta_5+
3\theta_6+\theta_7)^2
\end{eqnarray}
which is the metric for the orbifold $\BC^3/\BZ_3$.

The above analysis can be carried over to GLSMs with arbitrary
charges. Let us concentrate on the class of GLSMs with charges
\begin{equation}
Q = \left(1, n_2, -n_3, -n_4\right)
\end{equation}
where, without loss of generality we have taken the first charge to
be unity and we also assume that $n_4 > n_3 > n_2$, where the $n_i$ are
positive integers. This is an unstable conifold like singularity. 
Now take the case where $n_2$ acquires a large vev.  The sigma model 
metric becomes, 
\begin{equation}
ds^2 = \sum_{i=2}^4\left(d\rho_i\right)^2 + 
\frac{\rho_1^2}{n_2^2}d\left(n_2\theta_1 - \theta_2\right)^2+
\frac{\rho_3^2}{n_2^2}d\left(n_2\theta_3 + n_3\theta_2\right)^2+
\frac{\rho_4^2}{n_2^2}d\left(n_2\theta_4 + n_4\theta_2\right)^2
\label{gen1}
\end{equation}
This is recognised as the metric for the space $\BC^3/\BZ_{n_2}$ with
the GLSM charge matrix 
\begin{equation}
Q = \left(n_2 -1 , n_3, n_4, -n_2\right)
\label{gen2}
\end{equation}
in the sense that when the Fayet-Iliopoulos parameter of the latter
becomes very large, we recover the metric of eq. (\ref{gen1}). 
\footnote{One might convert this charge matrix to standard form by
making one of the integers to be unity using the fact that the integers
in eq. (\ref{gen2}) are defined modulo $n_2$, but that will not affect
the physics.} Now, we might add the twisted sectors corresponding to
this orbifold. E.g. enlarging the charge matrix by adding the first
twisted sector corresponding to eq. (\ref{gen2}), we obtain
\begin{equation}
Q = \pmatrix{1&n_2&-n_3&-n_4&0\cr n_2-1&0&n_3&n_4&-n_2}
\label{gen3}
\end{equation}
Now it is seen that making the first field in eq. (\ref{gen3}) very
large, we recover the $U(1)$ charge matrix
\begin{equation}
Q = \left(n_2^2 - n_2, -n_2n_3, -n_2n_4, n_2\right)
\label{gen4}
\end{equation}
From our assumption about the integers $n_i$, this is seen to be 
a non-supersymmetric unstable conifold like singularity as well. 
Thus we see that our Lagrangian analysis predicts the existence of
lower order conifold like singularities, which might be the decay
product of such a singularity of higher order.
It should be pointed out that in the above analysis, we need to take
care of the GSO projection of the twisted sectors. In general, a 
twisted sector will survive the type II GSO projection for
$\sum_i Q_i = {\rm even}$. For the purpose of our analysis, we will
broadly consider type 0 strings, it being understood that for type II
theories, some of the twisted sectors are projected out. 

In the previous paragraph, we considered the field with a relative 
positive charge being given a very large vev. A similar analysis 
can be done with any other field. 
As a concrete example, consider the GLSM of four fields $\phi_i,~
i=1\cdots 4$, given by the charge matrix
\begin{equation}
Q = \left(1,3,-5,-11\right)
\label{gen11}
\end{equation}
The Lagrangian for this model is as usual given by setting $m=4$ in
eq. (\ref{onepar}). We consider the various limits of the model by 
setting one field large at a time. Setting the vev of $\phi_1$ to
be very large, we recover flat space. A similar analysis for the
Lagrangian with $\phi_i, i = 2,3,4$ gives rise to the sigma model
metrics
\begin{eqnarray}
ds_2^2 &=& d\rho_1^2 + d\rho_3^2 + d\rho_4^2 
+ \frac{\rho_1^2}{9}
\left(3d\theta_1 + 2d\theta_2\right)^2 + \frac{\rho_3^2}{9}
\left(3d\theta_3 + 2d\theta_2\right)^2 \nonumber\\
&+& \frac{\rho_4^2}{9}
\left(3d\theta_4 + 2d\theta_2\right)^2 \nonumber\\
ds_3^2 &=&  d\rho_1^2 + d\rho_2^2 + d\rho_4^2 
+ \frac{\rho_1^2}{25}
\left(d\theta_3 + 5d\theta_1\right)^2 + \frac{\rho_2^2}{25}
\left(3d\theta_3 + 5d\theta_2\right)^2 \nonumber\\
&+& \frac{\rho_4^2}{25}
\left(4d\theta_3 + 5d\theta_4\right)^2\nonumber\\
ds_4^2 &=& d\rho_1^2 + d\rho_2^2 + d\rho_3^2 
+ \frac{\rho_1^2}
{\left(11\right)^2}
\left(d\theta_4 + 11d\theta_1\right)^2 \nonumber\\
&+& \frac{\rho_2^2}{\left(11\right)^2}
\left(3d\theta_4 + 11d\theta_2\right)^2+ \frac{\rho_2^2}{\left(11\right)^2}
\left(6d\theta_4 + 11d\theta_3\right)^2
\end{eqnarray}
where the subscripts on the r.h.s indicates which field has been made
large. These metrics are recognised to be the sigma model metrics for 
the orbifolds $\BC^3/\BZ_{3(2,2,2)}$ (or, equivalently, the supersymmetric
$\BC^3/\BZ_{3(1,1,1)}$), $\BC^3/\BZ_{5(1,3,4)}$ and 
$\BC^3/\BZ_{11(1,3,6)}$ orbifolds respectively. It is now clear how to 
enlarge the charge matrix. Including the relevant (and marginal) 
twisted sector states gives the enlarged charge matrix
\begin{equation}
Q = \pmatrix{1&3&-5&-11&0&0&0&0\cr1&0&1&1&-3&0&0&0\cr
1&3&6&0&0&-11&0&0\cr 2&6&1&0&0&0&-11&0\cr
4&1&2&0&0&0&0&-11}
\end{equation} 
and it is seen that apart from the third row, all other entries 
survive the type II GSO projection. Using our Lagrangian formulation,
we can analyse the phases of this theory in full generality. 
E.g taking the truncated charge matrix (corresponding to the first
two entries)
\begin{equation}
Q = \pmatrix{1&3&-5&-11&0\cr 1&0&1&1&-3}
\end{equation}
we see that making the first field very large, we get the GLSM
(corresponding to an unbroken $U(1)$) with charge matrix
$Q=\left(3,-6,-12,3\right)$. Similarly, assigning a very large vev to 
the third field gives an unstable $\BZ_{16}$ orbifold of $\BC^3$ etc.

The analysis with the full charge matrix is also simple. In this 
example, we get $42$ distinct phases of the full GLSM, $22$ of which
corresponds to flat space, one each of $\BZ_{11}$ and $\BZ_5$ orbifolds, 
$2$ each of $\BZ_4$ and $\BZ_6$ orbifolds, $5$ are $\BZ_3$ and
$9$ are $\BZ_2$ orbifolds. The exact action of these orbifolds 
can also be determined using the relevant Lagrangian in these phases. 
This illustrates the computational simplicity of our method of
determining phases of generic GLSMs.

\section{Conclusions}

In this paper, we have extended the analysis of \cite{ts1}
to study the phases of arbitrary GLSMs using the Lagrangian formulation.
Our analysis gives a simple and powerful way of obtaining these phases,
by tuning the fields which appear in the GLSM. We have shown how to 
construct the full phase structure of arbitrary charged GLSMs, which 
might be unstable. To us, this completes the analysis initiated 
in \cite{ts1}, and our results are complementary to those obtained
in \cite{narayan1},\cite{mn}. However, there are certain issues that 
need to be examined.
 
As we have indicated, the behaviour of D-branes in various phases of
these GLSMs can also be analysed using our formalism, even when the models
in question do not have an SCFT description. Consider, e.g the 
behavior of the world sheet gauge fields in the NLSM limit. For the
two parameter example, the expressions for the gauge fields become
\begin{eqnarray}
V^1_{\mu} = - \frac{\sum_{j_1,j_2}
\Delta(j_1,j_2) (K_{1j_1} Q^2_{j_2} \left|\phi_{j_2}\right|^2 - 
K_{1j_2} Q^2_{j_1} \left|\phi_{j_1}\right|^2 )^2} 
{\sum_{j_1,j_2} \left|\phi_{j_1}\right|^2 \left|\phi_{j_2}\right|^2 
\Delta(j_1,j_2)^2}\nonumber\\
V^2_{\mu} = - \frac{\sum_{j_1,j_2} \Delta(j_1,j_2) (K_{1j_2} Q^1_{j_2} 
\left|\phi_{j_1}\right|^2 - K_{1j_1} Q^2_{j_2} 
\left|\phi_{j_2}\right|^2 )^2} {\sum_{j_1,j_2} \left|\phi_{j_1}\right|^2 
\left|\phi_{j_2}\right|^2 \Delta(j_1,j_2)^2}
\end{eqnarray}
where
\begin{equation}
K_{1i}= \frac{i}{2}(\bar{\phi_i} \partial_1 \phi_i - 
\phi_i \partial_1\bar{\phi_i})
\end{equation}
These equations can be used to study the evolution of the D-brane 
boundary conditions for unstable spaces. As a simple example, the
open string GLSM boundary condition $D_1\phi_i=0$ (with the world 
sheet gauge field strength $v_{01}=0$) \cite{gjs} for the simplest 
fractional D-2 brane (other fractional D-2 branes are related to 
this by a quantum symmetry) in the unresolved orbifold phase
can be seen from these formulae to translate into the simpler 
condition $\partial_1\phi_i=0$ where the angular part of the $\phi_i$ 
now correspond to a gauge invariant angle. A similar result is obtained
for other phases as well. In \cite{mt}, the space time gauge field 
was computed for the $\BC/\BZ_n$ orbifolds and the behaviour of D-branes 
under the decay of this orbifold was studied by looking at the behavior
of this field with the decay of the singularity. It would be very 
interesting to do the corresponding analysis for generic unstable spaces 
using our Lagrangian formalism, especially for cases where there might
be terminal singularities. 

Further, having completely studied the full phase structure of a given
GLSM, one might ask if the reverse engineering of singular spaces is
possible. That is, given a certain number of orbifold singularities,
is it possible to construct a GLSM that will have these orbifolds in
their phases. An answer to this question will probably help us to 
have a better understanding of the D-brane quiver gauge theories 
corresponding to arbitrary GLSMs, that have been analysed in the 
supersymmetric case in \cite{labctoric}.

\end{document}